\documentclass[aps,%
reprint,
prl,
groupedaddress]{revtex4-2}
\usepackage{color}
\usepackage{hyperref}
\usepackage{float}
\usepackage{amsmath,amsfonts,amssymb,epsfig,graphicx}
\usepackage{slashed}

\newcommand{\jpsi}{J/\psi}

\begin{document}
\title{Charged lepton flavor violation searches in the charmonium system}

\author{Qiang \surname{Li}}
\email[]{qliphy0@pku.edu.cn}
\affiliation{State Key Laboratory of Nuclear Physics and Technology, School of Physics, Peking University, Beijing, 100871, China}

\maketitle

In the Standard Model (SM), the three active neutrinos are found to oscillate and mix among themselves. However, flavor mixing among their cousins, i.e., charged leptons, has never been observed and remains a mystery to be explored. 

Due to the Glashow–Iliopoulos–Maiani mechanism, the event rates of various Charged Lepton Flavor Violation (CLFV) processes, such as $\mu\rightarrow e\gamma$, $\mu\rightarrow 3e$ and $\jpsi\to e^{\pm}\mu^{\mp}$, are too tiny to be observed~\cite{Workman:2022ynf}, if the only CLFV source is from the neutrino side. Nevertheless, many models beyond the SM (BSM) predict CLFV effects can be largely enhanced. For example, models such as the GUTs (grand unified theories) and SUSY (supersymmetry) predict $\jpsi\to e^{\pm}\mu^{\mp}$ decay rates up to a detectable level of around $10^{-8} \sim 10^{-16}$ . The observation of any CLFV process thus would serve as a clear signal of new physics beyond the SM.

Indeed, the search for CLFV has attracted active interest since it has the potential to probe new physics indirectly at energy scales much higher than what is going to be accessible by the colliders in the foreseeable future~\cite{Calibbi:2017uvl}. Various searches (detailed in ref.~\cite{Workman:2022ynf}) have been performed at the muon-based experiments, B factories, and high energy colliders such as LEP and LHC, targeting the CLFV decays of $\mu$, $\tau$, Z and Higgs bosons. Interesting limits on CLFV are obtained in meson decays as well, with motivations mentioned as above. 

In the charmonium sector, the BES experiment has its clear advantage. Previously, it has found that $\mathcal{B}(\jpsi\to \mu^{\pm}\tau^{\mp})<2.0\times10^{-6}$ by analyzing $58\times10^{6}~\jpsi$ events~\cite{BES:2004jiw}, together with $\mathcal{B}(\jpsi\to e^{\pm}\mu^{\mp})<1.6\times10^{-7}$~\cite{BESIII:2013jau} and $\mathcal{B}(\jpsi\to e^{\pm}\tau^{\mp})<7.1\times10^{-8}$~\cite{BESIII:2021slj}. 

An updated search has been performed for the lepton flavor violating decay $\jpsi\to e^{\pm}\mu^{\mp}$, using $8.998\times 10^{9} \jpsi$ events collected with the BESIII detector at the BEPCII $e^+e^-$ storage ring. The upper limit of the branching fraction is determined to be $\mathcal{B}(\jpsi\to e^\pm\mu^\mp)<4.5\times10^{-9}$ at the $90\%$ C.L~\cite{BESIII:2022exh}, which improves the previous limit by a factor of more than 30. It is currently the most stringent limit on CLFV in heavy quarkonium systems and provides constraints on the parameter spaces of new physics models.


\bibliographystyle{ieeetr}
\bibliography{h}
\end{document}